\documentclass[epj]{svjour}

\usepackage{graphics}
\usepackage{courier}
\usepackage{cite}
\usepackage{amsmath}  
\usepackage{dirtree}
\usepackage{orcidlink}

\definecolor{codegray}{rgb}{0.9, 0.9, 0.9}
\newcommand{\texttg}[1]{\texttt{\colorbox{codegray}{#1}}}

\begin{document}
\title{Event-Generator Validation with MCPLOTS and LHC@home}
\author{
N. Korneeva\inst{1,2}\thanks{natalia.korneeva@cern.ch}\orcidlink{0000-0003-2461-6419} \and
A. Karneyeu\inst{3}\orcidlink{0000-0001-9983-1004} \and
P. Skands\inst{2,4}\orcidlink{0000-0003-0024-3822}
}
\institute{Tomsk Polytechnic University, 634050 Tomsk, Russia \and 
School of Physics and Astronomy, Monash University, Clayton VIC-3800, Australia \and Institute for Nuclear Research, 117312 Moscow, Russia \and Rudolf Peierls Centre for Theoretical Physics, University of Oxford, Parks Road, Oxford, UK}
\date{}
%
\abstract{
We document several recent updates to the MCPLOTS event-generator validation resource
. 
The project is based on the RIVET analysis library and harnesses volunteer computing provided by LHC@home to generate high-statistics MC comparisons to data. Users interact with the resource via a simple website, \url{mcplots.cern.ch}, which provides flexible options for requesting comparison plots and comprehensive statistical analyses on demand, all in a few clicks. 
The project has been structured to enable community-driven developments, and we discuss the computational back end, the web front end, and how to add new data analyses, generators, and tunes that would be accessible on the website for comparison. 
%
} 
\maketitle
\section{Introduction}
\label{intro} 

In particle physics, Monte Carlo event generators (MCs) connect theoretical calculations with the complex final states that are observed in experiments. For the experimental community, MCs serve as benchmarks for establishing calibrations and uncertainties, and for optimising measurements and detector designs.  For the theoretical community, they serve as platforms for exploring new approaches to solving perturbative quantum field theories, modelling their non-perturbative aspects, and exploring the observable consequences of physics beyond the Standard Model. 


When new experimental measurements are published, the associated analysis papers typically include comparisons to  small sets of representative MC models. These give an instantaneous snapshot of the theoretical state of the art at the time the analysis was done but can necessarily neither be totally exhaustive nor do they remain up to date as further theoretical work is developed and published. 

Major steps towards ensuring that experimental measurements remain useful to constrain theoretical models were the developments of the data preservation resource HEPDATA\footnote{\url{https://www.hepdata.net}}~\cite{Whalley:1989mt,Buckley:2006np,Maguire:2017ypu}, analysis preservation tools like HZTOOL~\cite{Waugh:2006ip} and RIVET~\cite{Buckley:2010ar,Bierlich:2019rhm}, suites of related analysis recasting and fitting tools~\cite{Buckley:2009bj,Cranmer:2010hk,Conte:2012fm,Dercks:2016npn,GAMBIT:2017yxo,Bellm:2019owc,Buckley:2021neu,Krishnamoorthy:2021nwv}, and the adoption of these tools by the physics community.  
These tools make it possible to validate new and alternative MC models in a homogeneous and standardised way. They also play an important role in the context of the growing field of MC tuning; see, e.g.,~\cite{Buckley:2009bj,Skands:2010ak,Schulz:2011qy,DeAlmeidaDias:2011wxe,Firdous:2013noa,Skands:2014pea,Yang:2014vra,Reichelt:2017hts,Kile:2017ryy,Brooks:2018tgf,Bellm:2019owc,Andreassen:2019nnm,Wang:2021gdl,Jueid:2023vrb,LaCagnina:2023yvi,Firdous:2023kye}. 

MCPLOTS builds on these tools to provide a quick-reference library for
validations of widely used MC models (and tunes) against a broad set
of reference measurements as well as a repository of suitable
generator run cards for each validation analysis. This is intended to
be useful for experts and non-experts alike. Even for experts, the
time required to generate statistically relevant event samples means
that it is rarely possible to get immediate feedback on a validation
question within the context of a single physics discussion. This can
be valuable, e.g., when plots from a paper are presented in a meeting,
and questions arise as to whether some new model can describe the
presented results better. Moreover, broad overviews involving many
different analyses and generator/tune combinations can require
significant time (and resource) investments.  

In general, \emph{any} of the several steps that are involved in producing a validation of a given MC model or tune against a specific set of experimental analyses can be time consuming and/or error prone, especially for (but not limited to) non-experts. For RIVET, which is the most widely used analysis preservation tool today and which also MCPLOTS is based on, these steps essentially comprise:
\begin{enumerate}
\item If starting from scratch: installing RIVET and the relevant MC generator(s), and any required dependencies.
\item Selecting a subset of RIVET analyses to include in the validation. 
\item \label{item:card} Preparing a run card for the given MC generator(s), for each selected RIVET analysis. 
\item \label{item:evt} Generating a statistically relevant number of events for each analysis.
\item \label{item:anl} Quantifying and analysing the agreement or disagreement, e.g., by computing and ranking some measure of statistical compatibility, and/or by simply making plots and organising them for inspection.
\end{enumerate}

For example, preparing run cards (step \ref{item:card}) requires knowledge of how to set up the specific MC(s) to generate the desired type of events correctly and efficiently. Such cards were not so far systematically available, and even experts have limits in terms of which generators and/or setups they are most familiar with. Another example of a task encountered frequently in MC generation applications, but which so far has received little dedicated attention, is the problem of determining an optimal set of generator-level phase-space cuts. We present two generic methods for this in sec.~\ref{sec:dev}. We note that we also make the run cards we use easily downloadable from the project website, which thus acts as a sort of repository of examples of such cards.

Thus, the aim with MCPLOTS is to provide an instantly accessible and broad set of comparisons that can be easily updated as new tunes, new generators (and new generator versions) are developed. The computing power is delivered by CERN's LHC@home volunteer computing project Test4Theory~\cite{LombranaGonzalez:2012gd,Karneyeu:2013aha,Barranco:2017gyv,Cameron:2019iyu}, and the original development of this project entailed significant innovations as Test4Theory was the first volunteer cloud in the world to make use of virtualisation~\cite{LombranaGonzalez:2012gd}. 

\begin{figure}[t]
        \centering
        \resizebox{0.9\textwidth}{!}{
            \includegraphics{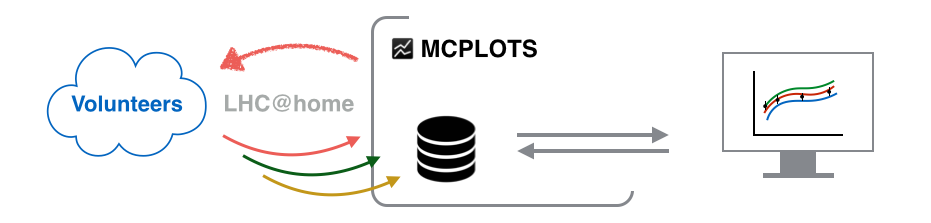}
        }
        \caption{The MCPLOTS workflow. Computing jobs are distributed to volunteers via the LHC@home platform. Completed jobs are stored on the MCPLOTS server, and descriptions of the obtained MC distributions - in the database, with the queries to which the website operates}
        \label{fig:mcplots-flow2}
\end{figure}

Likewise quantifying and analysing the level of agreement (step \ref{item:anl}) may seem trivial at face value, whether by producing a set of validation plots and inspecting them, or by performing an automated statistical analysis of them. But modern MC validation and tuning studies often involve \emph{very many} individual plots, and hence the question of organising and presenting an abundance of information becomes relevant in its own right. E.g., the current Test4Theory project runs over 100 RIVET analyses with over 100 different generator-version combinations, and the MCPLOTS site makes close to 4 million plots available within a few clicks. 
The overall workflow is schematically illustrated in Fig.~\ref{fig:mcplots-flow2}.

In this paper, we summarise how the MCPLOTS project addresses each of the steps 1--5 above, and how the project has evolved since its first publication~\cite{Karneyeu:2013aha}. We also present extensive advice and how-to information intended to assist community-driven additions and further development of the project.

\section{The Test4Theory Project and Volunteer Computing in MCPLOTS}
\label{sec:t4t}

The idea of creating an online repository of MC plots began around 2010\footnote{A precursor to this idea was the \textsc{JetWeb} project~\cite{Butterworth:2002ts,Buckley:2006np}, which among other things spurred the development of RIVET. The first sketches for what became \textsc{MCPLOTS} were made in connection with a Les Houches workshop in 2007~\cite{Skands:2007zz,Buttar:2008jx}.}.
By then, CERN was already running a successful volunteer computing project,  LHC@home, in which computing power donated by volunteers was harnessed (via the Berkeley Open Infrastructure for Network Computing, BOINC~\cite{Boinc19}) to run the SixTrack simulation of beam dynamics~\cite{Schmidt:1994kc}. Due to the heterogeneous nature of volunteer resources, SixTrack for LHC@home  had to be able to run natively on Windows operating systems, which represent majority of volunteer architectures.  SixTrack had accomplished this and was reaping the rewards in terms of access to tens of thousands of CPU cores donated by volunteers excited about contributing to CERN's scientific mission. But the requirement of native Windows applications was an impediment to easy adaptation of other scientific computing applications --- which are often developed exclusively with UNIX-compatible platforms in mind. 

To overcome this, the LHC@home developers had begun experimenting with Virtual Machines (VMs), based on the then nascent CernVM project~\cite{Buncic:2008zz}. The idea was that this would enable any application that could be built in a standardised scientific-computing environment to run on volunteer resources. However, BOINC had not been developed with this in mind, and significant development work had to be undertaken to pave the way for the world's first virtual volunteer cloud; see~\cite{LombranaGonzalez:2012gd,Barranco:2017gyv} for some of this history. Most of this work came from a small team spanning CERN's IT and PH divisions. 

By 2010, this team was looking for good test applications. Since software stacks in experimental HEP are typically quite deep, with comparatively large and complex footprints, the team  also reached out to CERN's Theory (TH) Division, where one of us had recently arrived as a new staff member. The Pythia event generator~\cite{Sjostrand:2006za,Sjostrand:2007gs} seemed well suited as a simple test case, with a quite small footprint and a simple build procedure being sufficient to do simulations that would have real physics interest. A collaboration was initiated, with the initial main goal of demonstrating whether realistic scientific applications could be run on volunteer architectures with the help of virtualisation. Since the Theory Division was involved, this pilot project got the operational name Test4Theory. 

Auspiciously, the first versions of the RIVET analysis toolkit had also recently been published~\cite{Waugh:2006ip,Buckley:2010ar}, and hence the scientific project became to combine LHC@home, CernVM, Pythia (and eventually other HEP MC generators), and RIVET. The addition of RIVET brought in the aspect of MC validation, making the name Test4Theory doubly appropriate, and hence this name stuck. 

\begin{figure}[tp]
    \centering
    \resizebox{\textwidth}{!}{
    \includegraphics{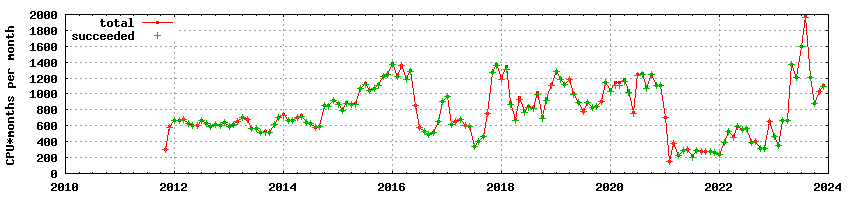}
    }
    \caption{Monthly average estimates of the instantaneous number of CPU cores connected to the Test4Theory project since logging began\label{fig:cpu}. Due to the heterogeneous and non-trivial setup of volunteer cloud resources, this number does not translate directly to a specific CPU benchmark but only gives an indication of the resources available and the evolution of this resource base with time. }
\end{figure}
By 2011, the first editions of the project website had been set up and alpha tests of using BOINC for MC production had started. The plot in Fig.~\ref{fig:cpu} shows the CPU resources accessed by the Test4Theory project since logging began in late 2011. Initially, Test4Theory was the only project on VirtualLHC@home and hence received 100\% of all CPU resources. This was the case until about 2016, during which the main CERN experiments CMS, ATLAS, and LHCb also joined the VirtualLHC@home platform. This led to a relative drop in the CPU resources allocated to Test4Theory, as can also be seen in the figure. Still, the volunteers base continued to grow and the drop was recovered by about 2020. Next, in 2021, VirtualLHC@home was unified with the primary LHC@home, to consolidate volunteer-based resources at CERN. This again led to a drop in the relative Test4Theory CPU share. By today, 2024, also this CPU drop has been fully recovered, with an average equivalent to about 1,000 CPUs contributing to the project at any given time.

\begin{figure}[tp]
    \centering
    \resizebox{0.8\textwidth}{!}{
    \includegraphics{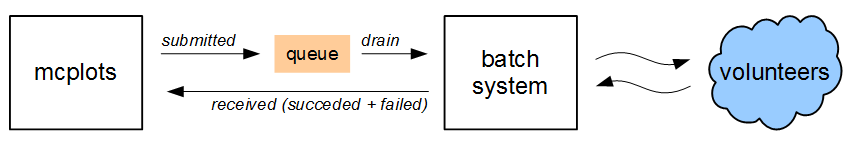}
    }
    \caption{The overview of jobs flow from MCPLOTS to volunteers and back. \label{fig:jobflow}}
\end{figure}

 Fig.~\ref{fig:jobflow} demonstrates the jobs flow in the ``MCPLOTS - LHC@home - volunteers'' system. The flow consists of the following steps:

\begin{enumerate}
  \item MCPLOTS generates jobs in batches, each of which can be as high as 100,000 jobs corresponding to the various combinations of physics process and physics model. Each job is a regular executable file containing all necessary scripts and programs to perform the simulation.
    
  \item Each job is submitted to the input queue of the LHC@home project (batch system).

  \item The batch system distributes the job executable to volunteers.

  \item Each job executable runs on a volunteer machine. Once that job is finished, all output files are packaged into a single output file.

  \item The output file is copied back to the batch system.

  \item The batch system collects the output files and moves these files to MCPLOTS.

  \item MCPLOTS process the output files to populate its histograms database and accumulate statistics.
\end{enumerate}

We also note that, apart from the raw CPU power, the LHC@home community has also provided quite a lot of useful feedback on the behaviour of the scientific applications, chiefly through the message boards\footnote{\colorbox{codegray}{\url{https://lhcathome.cern.ch/lhcathome/}}}. This can be, for instance, reports of software crashes or unusual progress log output. This feedback has been very valuable for Test4Theory/MCPLOTS as there are hundreds of thousands of combinations of job configurations, which are not always possible to fully debug in advance before the submission to the LHC@home system. Hence a small group of quite active ``expert'' volunteers have effectively also provided a human level of control on the otherwise fully automated jobs flow, in many cases leading to the identification (and flagging/removal) of particularly buggy generator versions, inappropriate generator/process setups, and the like.

\section{The Project}
\label{sec:project}
    
 MCPLOTS is a simple browsable repository of plots comparing high-energy physics event generators to a wide variety of available experimental data, for tuning and reference purposes. Apart from individual plots contained in papers and presentations, there has not previously been any central database where people can quickly see how tune X of version Y of generator Z looks on distribution D. The idea with MCPLOTS is to provide such a repository, based on:
\begin{itemize}
        \item The RIVET analysis tool~\cite{Buckley:2010ar,Bierlich:2019rhm}.
        \item MC event generators~\cite{Buckley:2011ms,Campbell:2022qmc};
        \item The LHC@HOME platform~\cite{LombranaGonzalez:2012gd,Karneyeu:2013aha,Barranco:2017gyv,Cameron:2019iyu}. 
\end{itemize}

Under the hood, this further crucially relies on the HEPDATA repository~\cite{Whalley:1989mt,Maguire:2017ypu}, the CERNVM File System~\cite{Blomer:2012cm,Blomer:2013tqa}, the BOINC platform for volunteer computing\footnote{\colorbox{codegray}{\url{http://boinc.berkeley.edu/}}}~\cite{Hoimyr:2012ur,Barranco:2017gyv,Boinc19}, and the regularly updated installations of HEP MC event generators on CERNVM-FS provided by the CERN EP-SFT group\footnote{\colorbox{codegray}{\url{https://ep-dep-sft.web.cern.ch/}}}. It of course also relies on the HEP MC event generators themselves\footnote{At present writing, these include, alphabetically, ALPGEN~\cite{Mangano:2002ea}, AMC@NLO~\cite{Alwall:2014hca}, EPOS LHC~\cite{Pierog:2013ria}, HERWIG++~\cite{Bahr:2008pv}, HERWIG~7~\cite{Bellm:2015jjp}, MADGRAPH~5~\cite{Alwall:2011uj}, POWHEG~\cite{Alioli:2010xd,Platzer:2011bc}, Pythia~6~\cite{Sjostrand:2006za}, Pythia~8~\cite{Bierlich:2022pfr}, SHERPA~\cite{Sherpa:2019gpd}, and VINCIA~\cite{Fischer:2016vfv}.} and on all of the experimental analyses that have been performed and implemented in RIVET. 

\begin{figure}[t]
        \centering
        \resizebox{0.95\textwidth}{!}{
            \includegraphics{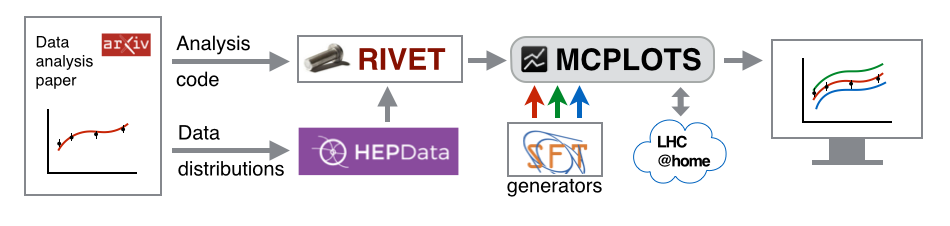}
        }
        \caption{The place of MCPLOTS in HEP MC validation workflows.}
        \label{fig:mcplots-flow}
\end{figure}

Fig.~\ref{fig:mcplots-flow} illustrates the role of MCPLOTS in modern HEP. The preservation of data points and data analysis codes are, by now, almost standard for modern experimental HEP publications, at least where applications to MC validation is an active consideration. Data points plotted in papers are published in the HEPData repository, and are also integrated into the RIVET repository along with corresponding analysis codes. The Test4Theory/MCPLOTS project runs the RIVET analyses, using MC generator installations provided by the CERN EP-SFT group. Multiple versions of each generator can be run, and each generator can be run for multiple different tunes. Which RIVET analyses, generator versions, and tunes to run are defined in the MCPLOTS configuration files (see below), which are periodically updated to include further RIVET analyses, new generators and versions, and new tunes (or other non-default generator options of interest). The resulting MC distributions are collected on the MCPLOTS server and can be accessed through the website. 

In terms of the internal architecture of the project, it is structured as a directory tree, with separate subdirectories for different purposes. The directory structure that can be downloaded from the repository \\ \texttg{https://gitlab.cern.ch/MCPLOTS/mcplots} can be represented as following:
\medskip
\dirtree{%
        .1 mcplots.
        .2 doc.
        .2 plotter.
        .2 scripts.
        .3 mcprod.
        .2 www.
}
\medskip
The \texttg{doc} directory contains up-to-date documentation that updates and extends this paper. Future developers are highly encouraged to consult this directory. The \texttg{plotter} directory contains the source code for the utility of the same name (\texttt{plotter.cc}) which is used to generate plots for the website and to calculate the $\chi^2$ value for each generated MC distribution for comparison with data. The C++ source code of the plotter only depends on ROOT and takes histogram/data files in the format used on the MCPLOTS site as input. It can be copied from the repository and modified for personal use, and several examples are included in the directory to test its behaviour on both well-formed and inconsistent/erroneous input. 

All scripts used to update the contents of the website are located in the \texttg{scripts} directory, and those used to organise and run generator jobs are in \texttg{scripts/mcprod}.  

Finally, all HTML and PHP source codes, and all style and configuration files required for the website to work are located in the \texttg{www} directory. 
    
Apart from the code repository, the experimental data points and the corresponding MC distributions produced by Test4Theory are stored in the \texttg{www/dat} directory. In previous versions of MCPLOTS~\cite{Karneyeu:2013aha}, $\chi^2$ values (defined as in Sec.~\ref{sec:comparison} below) for each MC distribution were calculated on-the-fly via user requests, but this gradually became prohibitively slow as the number of required calculations increased. Therefore, $\chi^2$ values are now re-computed and stored centrally each time the underlying data is updated. This allowed the $\chi^2$ comparison functionality on the site to speed up significantly. The tabulated $\chi^2$ values are stored in the \texttg{www/dat/cache/chi2} directory, and are organised in text files, each of which corresponds to a separate hard process for a specific generator-version-tune combination, and each of these files contains the $\chi^2$ values for all relevant MC distributions.
    
Thus, $\chi^2$ values corresponding, say, to the ``jets'' process type in proton-proton and proton-antiproton collisions, generated with the Pythia~8 event generator, version 8.244, tune \texttt{tune-2m} are stored in the file named
    
\medskip
\noindent\texttt{pythia8-8.244-tune-2m-ppppbar-jets.txt}. 
    
\medskip
\noindent Each of the above-mentioned distributions is described in this file by a line like the following
    
\medskip
    
\noindent\texttt{\color{blue}pythia8-8.244-tune-2m\color{black}-{\color{orange}ppppbar-jets}--{\color{magenta}js\_int--1960--cdf3-037\\--CDF\_2005\_S6217184--d07-x01-y01} {\color{cyan}0.73} \color{red}90 91},
    
\medskip
\noindent which consists of the {\color{blue}generator-version-tune} (G-V-T) combination and the {\color{orange}hard process}, some {\color{magenta}MCPLOTS-internal information} which will be described in more detail later in this section (variable name \texttt{js\_int}, CM energy \texttt{1960}, cuts \texttt{cdf3-037}, and RIVET reference \texttt{CDF\_2005\_S6217184--d07-x01-y01}), the {\color{cyan}$\chi^2$} value itself, and the (MCPLOTS-internal) IDs of the {\color{red}MC} and the corresponding {\color{red}data} distribution. Note: the colour highlighting here is just for emphasis and is not present in the text file. 
    
This storage organisation allows to quickly calculate the mean $\chi^2$ value on a given hard process for a particular G-V-T combination and present this data on the website in the form of tables based on thousands of distributions, as will be discussed in detail in Sec.~\ref{sec:comparison}.

\begin{figure}[t]
\centering\noindent
\scalebox{1.01}{\ttfamily\scriptsize
        \noindent
        \begin{tabular*}{\linewidth}{l@{\extracolsep{3pt}}l@{\extracolsep{8pt}}l@{\extracolsep{0pt}}lllllll@{\extracolsep{4pt}}l@{\extracolsep{4pt}}l@{\extracolsep{4pt}}l@{\extracolsep{8pt}}l}
            \texttt{[id]}&[fname]&[type]&[prc]&[obs]&[tune]&[exp]&[ref]& [hid]&[beam]& [Ecm]&[cuts]&[gen]&[ver]\\
            {\color{red}90}& dat/CDF\_2005& mc &jets & js\_int & tune-2m & CDF&CDF\_2005&d07& ppbar & 1960&cdf3&pythia8&8.244\\
            & \_S6217184-ppbar&  & & & & &\_S6217184&  -x01&& &-037&&\\
            &-1960/jets-js\_int &  &&  &  &  & & -y01 && &&& \\
            &-cdf3-037-d07-x01&  &&  &  &  & &  && &&& \\
            & -y01/pythia8 &  &&  &  &  & &  & &&&& \\
            & -8.244-tune-2m.dat &  &&  &  &  & &&  & &&& \\
            {\color{red}91}& dat/CDF\_2005 & data &jets& js\_int &  & CDF & CDF\_2005 &d07& ppbar & 1960&cdf3&& \\
            & \_S6217184-ppbar&  & & & & &\_S6217184&  -x01&& &-037&&\\
            & -1960/jets-js\_int &  &&  &  &&  &-y01 &  & &&& \\
            & -cdf3-037-d07-x01 &  &&  &  &  && &  & &&& \\
            & -y01/CDF\_2005 &  &&  &  &  & &  & &&&& \\
            & \_S6217184.dat &  &&  &  &  & &  & &&&& \\
        \end{tabular*}%
}
\caption{Example of the contents of the \texttg{histograms} table. \label{fig:histograms}}
\end{figure}
Information about all MC and data distributions on the server is stored in a database. It is a relational database  consisting of several tables, where the \texttg{histograms} table contains information about the distribution itself and the other tables contain technical information about production. All user interaction with the website is done through queries to the \texttg{histograms} table.

An example of the description of the above-mentioned MC and the corresponding data distributions in the \texttg{histograms} table is shown in Fig.~\ref{fig:histograms}. These particular two lines describe an integrated jet-shape distributions (the internal MCPLOTS name is encoded in the \texttt{[obs]} column): the first one corresponds to the MC distribution (with unique MCPLOTS ID 90, with colour emphasis as above) and the second one to the data distribution (with unique MCPLOTS ID 91). The type (MC or data) is made explicit in the third column, \texttt{[type]}. The second column shows the paths to the given distributions on the MCPLOTS server. The \texttt{[prc], [tune], [beam], [Ecm], [gen], [ver]} columns are the parameters of the distribution, and \texttt{[cuts]} provides additional information to be shown on the website. \texttt{[Ref]} is a RIVET analysis reference, where the distribution is taken from, \texttt{[hid]} is the internal RIVET ID of this distribution and \texttt{[exp]} is a corresponding experiment name. 
    
When the data is updated, the database update is performed by running the script, that scans through the generated distributions in the \texttg{dat} directory, extracts the necessary information from the file path, and adds it to the table.

\section{User's Guide: The MCPLOTS Website}
\label{sec:website}

\begin{figure}[tp]
\centering
\begin{minipage}[t]{0.8\textwidth}
    \resizebox{1\textwidth}{!}{
        \includegraphics{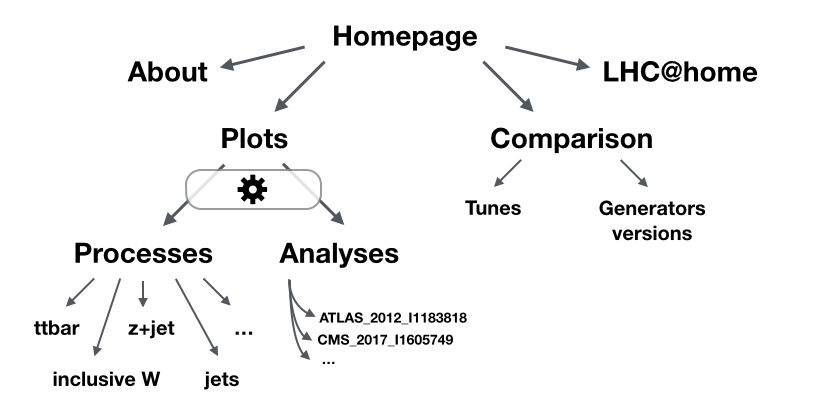}
        }
    \caption{MCPLOTS website structure}
    \label{fig:website}
\end{minipage}
\end{figure}
\begin{figure}[tp]\centering
\begin{minipage}[t]{0.8\textwidth}
    \resizebox{1\textwidth}{!}{
        \framebox{\includegraphics{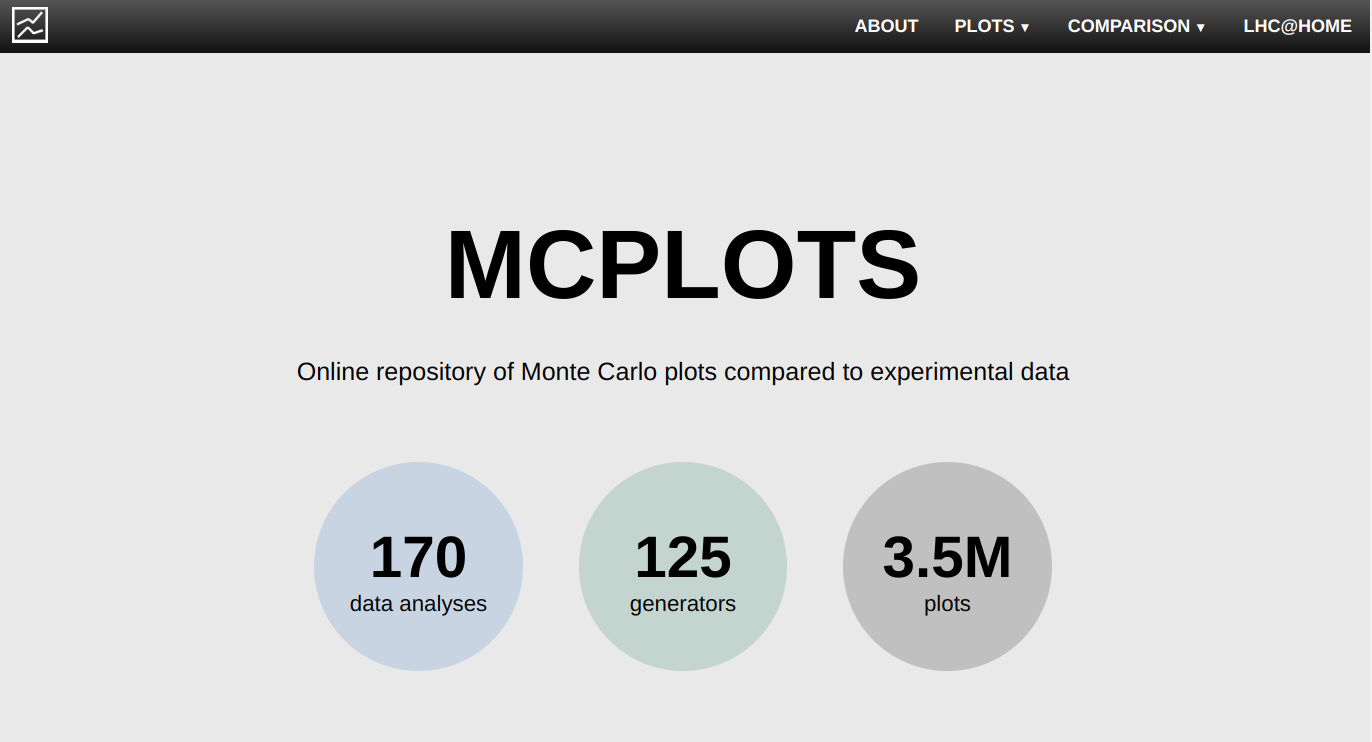}}
    }
    \caption{The main MCPLOTS website page}
    \label{fig:mainpage}
\end{minipage}
\end{figure}
    
The project website \colorbox{codegray}{\url{http://mcplots.cern.ch/}} has recently been redesigned and its current structure is shown in Fig.~\ref{fig:website}. An immediate visual change is that new version has abandoned the earlier ``all-in-one'' interface~\cite{Karneyeu:2013aha} in order not to overload the main menu and present the user with a more clean, modern interface. 

The front page, shown in Fig.~\ref{fig:mainpage}, contains links only to the principal project sections: \textbf{Plots} and \textbf{Numerical comparison}, as well as to the \textbf{About} page and an external link to the LHC@home project. These main sections can be accessed via the fixed top navigation menu with a drop-down menu containing the available options: hard processes for the \textbf{Plots} menu option and generator or tune tables for \textbf{Comparison}. Also, without selecting options from the drop-down menu, but by clicking on the section name, one can get to the corresponding section of the website, with customisation options and a brief description of the functionality at hand. Besides, the top navigation drop-downs have been removed in the smartphone version of the website.  
    
Without going into detail on the \textbf{About} section, which contains a lot of information about the project, user's and developer's guides, complementing the present Article, here is a description of the main features of the website.
    
\subsection{The PLOTS Section}

The \textbf{Plots} section is the main section of project that allows to compare the MC distributions generated within MCPLOTS with each other and with the data distributions. The main \textbf{Plots} page can be accessed by clicking on its name in the navigation menu, and it contains options to control and organise plots to be displayed.
    
    \begin{figure}[t]
        \centering
          \framebox{\includegraphics[width=0.85\textwidth]{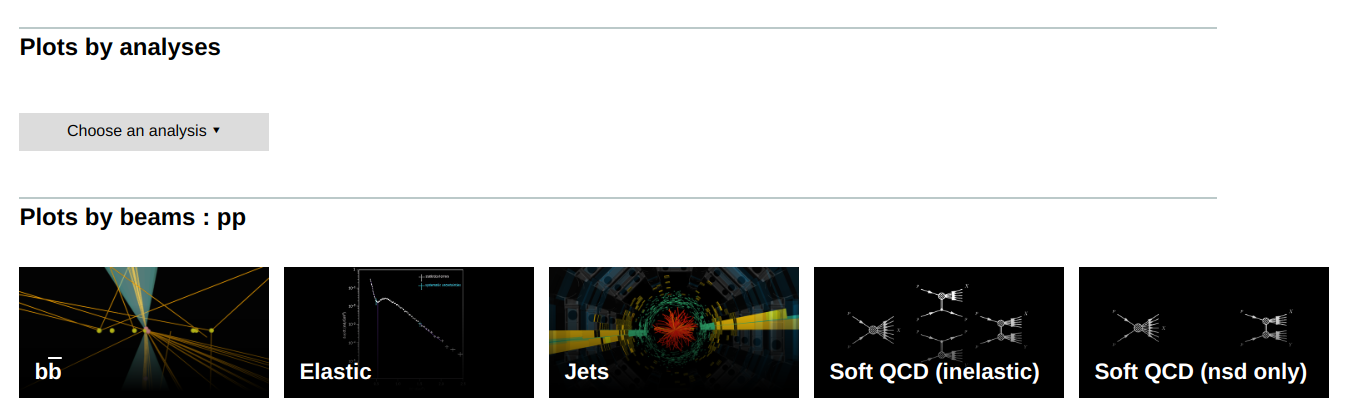}}
            \caption{Options to filter the plots by a RIVET reference or by a hard process.}
            \label{fig:anfilter}\vskip5mm
                \framebox{\includegraphics[width=0.85\textwidth]{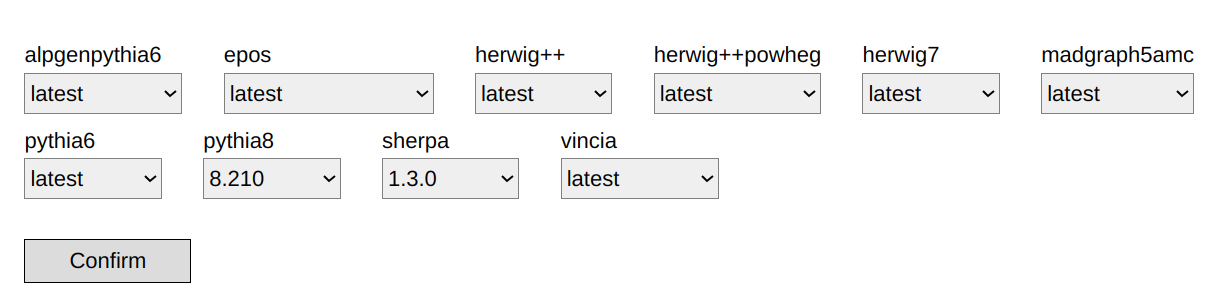}}
            \caption{The menu for selecting the default generator version. The default value is "latest".}
            \label{fig:genver}
    \end{figure}
    
The main possibilities, shown in Fig.~\ref{fig:anfilter}, are to filter the distributions to be displayed by either the hard process or the RIVET data analysis reference realised as a dropdown menu. The latter currently requires knowledge the RIVET ID of the analysis of interest. The ID is typically formed from the experiment name, the year, and the inSPIRE ID (or SPIRES ID, for older analyses) of the paper containing the original analysis; the references beginning with ``I'' are inSPIRE codes, while ones beginning with ``S'' are SPIRES ones).

Note: at the bottom of the page there is a menu for selecting generator versions. The default on MCPLOTS is always to display the most recent generator versions that are available on the site. But distributions generated with previous versions are also stored, and one can choose a specific generator version to be displayed on plots.  The menu for selecting specific versions is shown in Fig.~\ref{fig:genver}, where the Sherpa and Pythia versions have been changed from their defaults. This specific version will then be shown on plots instead of the latest one, and this selection remains in effect until the user leaves the \textbf{Plots} section, or until another change is made.

\begin{figure}[tp]
         \centering
                \framebox{\includegraphics[width=0.98\textwidth]{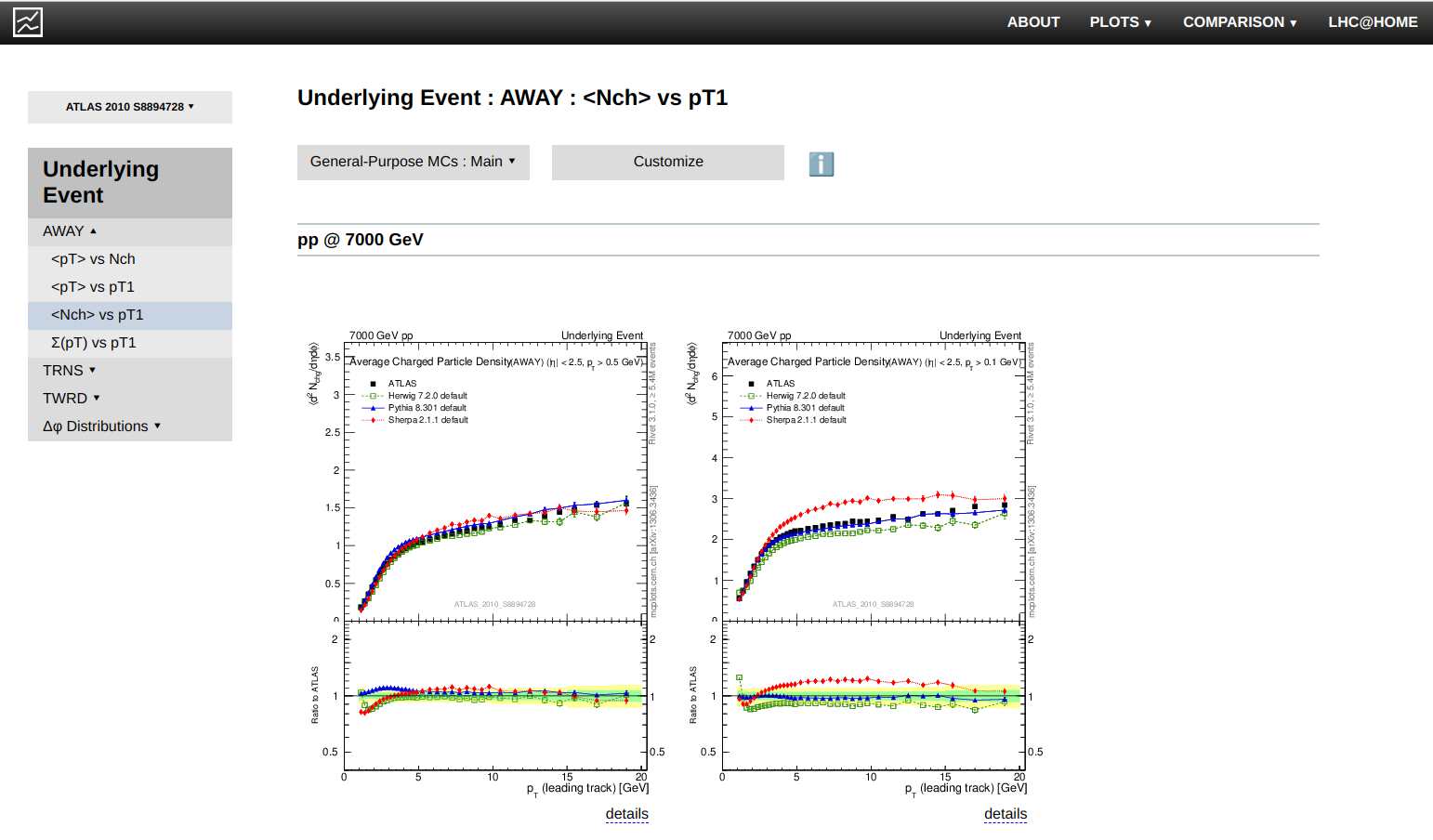}}
            \caption{Typical plots page. The left menu provides access to all available for a given hard process distributions, which appear in the main part of page.}
            \label{fig:plotspage}
            \end{figure}
            \begin{figure}[tp]\centering
                \framebox{\includegraphics[width=0.36\textwidth]{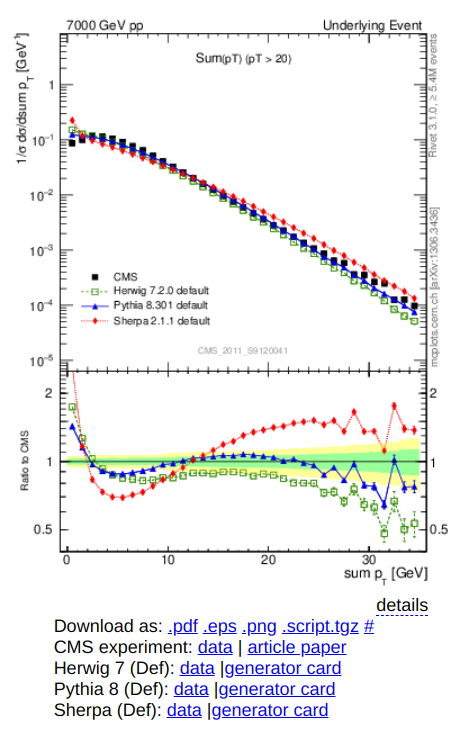}}
            \caption{Typical plot with the open dropdown "details" section.}
            \label{fig:typplot}
    \end{figure}
    
By selecting a data analysis or hard process from the options available, the user is taken to a dynamically generated page where the plots of the selected analysis/process will be shown. A typical view of such plots page is shown in Fig. \ref{fig:plotspage}. The left hard process menu provides access to all (represented in the selected data analysis, if applicable) available for a given process distributions, some of which are organised into groups. Usually only one hard process is considered in a data analysis, in cases where this is not the case, there will be additional hard process menus on the page. The data analysis selection can be changed using the drop-down menu just above the hard process menu; note that this analysis menu is not available if the hard process plots filter is selected. Finally, the hard process can be quickly changed via the top navigation menu.
    
When clicking on any observable in the left menu, the requested distributions, grouped by beam energy and starting from the highest one, appear in the main part of the page, with the default set of MC curves. 
    
At the top of the \textbf{Plots} pages, a dropdown menu allows to select between many different predefined sets of MCs (and tunes)  to be plotted. By default, the default tunes of the main general-purpose MCs are shown. Other predefined selections available through this dropdown menu currently include soft-inclusive MCs, Matched/Merged MCs, and more dedicated comparisons focusing on each of the main general-purpose generators. For Pythia in particular, this typically includes several different options for which tunes to display. 
If further combinations are required, 
a second ``Customize'' button allows the user to take full control of which MCs / tunes (among those that are available on the site) to include on the plots. 
 The selected preset applies to all plotted distributions and will remain in effect until the user leaves the \textbf{Plots} section or selects another preset. Clicking the ``Customize'' button opens a menu from which the user can select freely among the available generator models, including their versions and various tunes. The full customisation  only applies to the current page, and when another variable is requested, the default MC set (or a previously selected preset) will be shown again.
    
An example of a ``typical'' plot from MCPLOTS is shown in Fig.~\ref{fig:typplot}. It is always the case that black squares represent the experimental data, wich are labelled with the experiment name in the legend of the plot. Coloured dots connected by lines correspond to the requested MC distributions. They are labelled with the generator name, its version, and its tune. At the top of each plot, information about the given distribution is shown, usually the observable name, and the (data analysis) cuts that were used to obtain this distribution. Axis labels are taken from RIVET, where they are provided by the authors of the given analysis. Above each plot, the beam, its energy, and hard process used to generate the distribution are shown. To the right of the plot is a reference to the MCPLOTS paper, the RIVET version in which it was obtained, and the number of MC events. If these numbers are different for different MC curves shown on the plot, the smallest one is provided. Finally, at the bottom of the plot, the RIVET reference of the data analysis is given. 
    
Underneath each plot is shown a ratio pane with the MC results normalised to the data. The vertical range of the ratio plot is fixed  to [0.5, 2], with a bit of margin on either side. Note that the ratio pane uses a logarithmic axis. The central green and yellow bands corresponds to the experimental uncertainty at 1$\sigma$ and 2$\sigma$ levels respectively. 

Below each plot is the \texttg{details} dropdown section. Using the links from this section, one can download the plot in higher resolution and/or in vector graphics formats. Currently, \texttg{.pdf}, \texttg{.eps} and \texttg{.png} formats are available. The \texttg{article paper} link provides access to the article where this distribution was obtained. The possibility to download all tables of experimental and MC results used to draw the plot on the MCPLOTS website is also provided by clicking on the \texttg{data} links. One can download all these tables and a steering file for the MCPLOTS \texttg{plotter} utility by clicking on the \texttg{.script.tgz} link. Finally, the steering files for the MC generators used to produce this plot are also provided under the \texttg{generator card} links. Thus, all plots are fully reproducible.
    
\subsection{The COMPARISON Section}
\label{sec:comparison}
     
In addition to visualisations and comparisons by eye of the various MC plots, the MCPLOTS infrastructure also provides the possibility to numerically compare each generated distribution with the experimental data. For this purpose, a $\chi^2$ criterion is used, which is calculated using the following formula:
    
 \begin{equation}
   \chi^2=\frac{1}{N_{\mathrm{bins}}}\sum_{i=1}^{N_\mathrm{bins}}\frac{\big(\mathrm{MC}_i-\mathrm{Data_i}\big)^2}{\sigma^2_{\mathrm{Data,}i}+\sigma^2_{\mathrm{MC,}i}+\big(\epsilon_{\mathrm{MC}}\mathrm{MC}_i\big)^2},
   \label{chi2}
\end{equation}
where $\sigma_{\mathrm{Data},i}$ is the uncertainty on the experimental measurement (combined statistical and systematic) of bin number $i$, and $\sigma_{\mathrm{MC},i}$ is the (purely statistical) MC uncertainty in the same bin. By default, the generator predictions are assigned a flat $\epsilon_\mathrm{MC} = 5\%$ ``theory uncertainty''~\cite{Karneyeu:2013aha,Skands:2014pea}, in addition to the purely statistical MC uncertainty as a baseline sanity limit for the achievable theoretical accuracy with present-day MC models. (This modified $\chi^2$ measure is sometimes denoted $\chi^2_{5\%}$.) A few clear cases of GIGO\footnote{Garbage In, Garbage Out.} are excluded from the $\chi^2$ calculation, but some problematic cases remain. Thus, e.g., if a calculation returns a too small cross section for a dimensionful quantity, the corresponding $\chi^2$ value will be large, even though the shape of the distribution may be well described. One could argue about how this should be treated, how much uncertainty should be allowed for each observable, how to compare consistently across models/tunes with different numbers of generated events, whether it is reasonable to include observables that a given model is not supposed to describe, etc. These are questions that we do not believe can be meaningfully (or reliably) addressed by a fully automated site containing tens of thousands of model/observable combinations. Finally, the normalisation factor $1/N_\mathrm{bins}$ is used for all $\chi^2$ calculations, because from the MC and data histograms alone it is difficult to determine unambiguously whether the number of degrees of freedom for a given distribution is $N_\mathrm{bins}$ or $(N_\mathrm{bins}-1)$.
     
\begin{figure}[tp]
         \centering
                 \framebox{\includegraphics[width=0.75\textwidth]{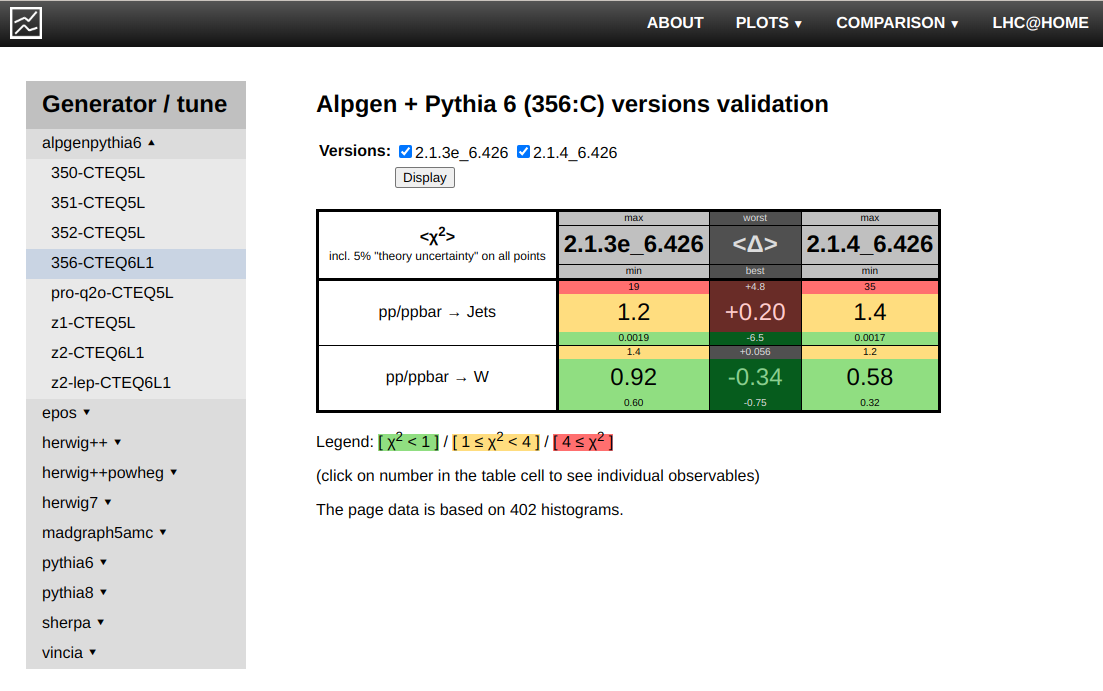}}
             \caption{Typical comparison page}
             \label{fig:typcomp}
\end{figure}      
     
There are two main comparison options on the site: 
\begin{itemize}
\item Compare different \textbf{Tunes} of an MC generator.
\item Compare different \textbf{Versions} of an MC generator. 
\end{itemize}
The former is intended to provide an overview of how different tunes of the same generator compare to each other. The latter is intended to help identify changes that have occurred between different generator versions and can help validate salient changes in the physics modelling or highlight  so-called ``version creep''. We so far do not provide options for analogous comparisons between different MC generators, as we believe there are substantial caveats in judging such comparisons, which would be difficult to account for properly in a fully automated setup like ours. 
    
The \textbf{Versions} option takes the user to a dynamically generated page showing a summary table of $\chi^2$ values corresponding to hard processes available for the selected generator. A typical view of such a page is shown in Fig.~\ref{fig:typcomp}. The menu on the left allows to select a generator and a specific tune, and above the table are shown the generator versions that are available for this selection. They are sorted in ascending order, i.e. earlier ones are shown first. One can select any of them to be shown in the table. 
     
This table aggregates the $\chi^2$ values averaged over all generated distributions for a given hard process (table rows) calculated according to Eq.~(\Ref{chi2}). This average value is shown in the centre of each table cell, while the maximum and minimum $\chi^2$ values for a given hard process, for the given tune, are shown at the top and bottom, respectively. The table uses colour differentiation of the presented $\chi^2$ values: for those less than 1 a green background is used, from 1 to 4 --- an orange one, and for values greater than 4 --- a red one, following the spirit of the Les Houches ``tune killing'' exercise. Under the table it is mentioned how many individual distributions were used in the calculation of the presented mean values. 
    
 \begin{figure}[tp]
 \centering
 \framebox{\includegraphics[width=0.75\textwidth]{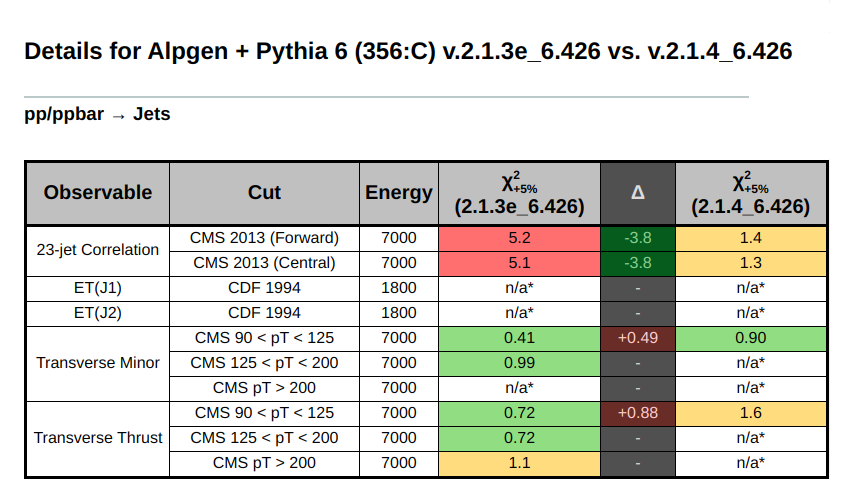}}
             \caption{Table of individual $\chi^2$ values.}
             \label{fig:typdet}
\end{figure}
The $\mathrm{\Delta}$ column between version columns shows the change in the mean $\chi^2$ value (in the centre of the cell), as well as maximum and minimum change (at the top and bottom, respectively).  Since the versions in the table are presented in chronological order, this column allows a quick estimation of changes in the average  $\chi^2$ value with the new releases. A similar colour scheme is used here: if the changes are minor, the $\mathrm{\Delta}$  values are shown on a dark grey background, if the $\chi^2$ value has decreased, the background is dark green, and if it has increased, the background is dark red.
     
The mean $\chi^2$ value in each cell is clickable and allows to see a table of $\chi^2$ values for each distribution individually, where the colour scheme described above is also used. The same goes for the main $\mathrm{\Delta}$ value which is also clickable and allows to see the same table of individual $\chi^2$ values, but for two versions at the same time, with the inset column of individual $\mathrm{\Delta}$s. An example of such a table is shown in the Fig. \ref{fig:typdet}. Additional information on the distributions in the Energy and Cuts columns can be found in this table.  The "n/a" value on the white background means that the MC distribution and the data distribution have different numbers of bins and the $\chi^2$ value calculation is not possible. All individual $\chi^2$ values, including those "n/a", are also clickable and allow to look at the corresponding plot. Clicking on an individual $\mathrm{\Delta}$ value will show the plot with 2 MC distributions corresponding to the selected versions, and the data distribution.
     
If one wants to compare different tunes of the generator, clicking on the \textbf{Tunes} option will bring up a page similar to the one shown in Fig. \ref{fig:typcomp}.  The left menu allows to select a generator and its specific version, and above the table all available for this selection tunes are listed, being sorted in alphabetical order.
     
A significant difference with the versions table is that in the absence of the need for a timeline comparison, there are no $\mathrm{\Delta}$ columns on the \textbf{Tunes} comparisons pages. 

\section{Developer's Guide}
\label{sec:dev}
    
Adding a new analysis, generator, or generator tune to the MCPLOTS project is done by adding the description of this analysis/generator/tune to the configuration files.  All such files are located in the directory:
\begin{center}
    \texttg{scripts/mcprod/configuration}
\end{center}
This of course assumes that the desired generator and RIVET version for desired analysis are available in the LCG releases software repository SFT located at:
\begin{center}
    \texttg{/cvmfs/sft.cern.ch}
\end{center}
The machinery provides sufficient flexibility to allow the usage of generator and RIVET versions from various sub-locations of the SFT repository.

\subsection{Adding a new RIVET analysis on MCPLOTS}
    
The implementation of a RIVET analysis in the MCPLOTS machinery is given by its description in the file \\ \texttg{rivet-histograms.map}. It contains information about all distributions used in the analysis: their description, internal MCPLOTS names and MC generator initialisation parameters to be used to produce the distributions.
    
The list of all distributions available in a given RIVET analysis are to be found in the corresponding RIVET \texttg{.plot} file, which contains the information needed to plot the distributions. Note that not all of them can be presented in the MCPLOTS repository: currently only one-dimensional distributions obtained with a fixed beam energy are supported.
    
As an illustration, consider the implementation of the recent CMS analysis \texttt{CMS\_2019\_I1764472}  \cite{CMS:2019fak}. The RIVET \texttg{.plot} file of this analysis contains the following information:
    
\begin{verbatim}
BEGIN PLOT /CMS_2019_I1764472/d01-x01-y01
Title=CMS, 13 TeV, jet mass in boosted top quark decays
XLabel=$m_\mathrm{jet}$
YLabel=$\frac{d\sigma}{dm_\mathrm{jet}} \frac{\mathrm{fb}}{\mathrm{GeV}}$
LogY=0
END PLOT

BEGIN PLOT /CMS_2019_I1764472/d02-x01-y01
Title=CMS, 13 TeV, normalized jet mass in boosted top quark decays
XLabel=$m_\mathrm{jet}$
YLabel=$\frac{1}{\sigma} \frac{d\sigma}{dm_\mathrm{jet}} \frac{1}{\mathrm{GeV}}$
LogY=0
END PLOT
\end{verbatim}
    
\noindent The two \texttt{BEGIN PLOT} $\ldots$ \texttt{END PLOT} sections indicate that this analysis contains 2 distributions: first, the $t\bar{t}$ differential cross section as a function of the jet mass, and second, its normalised version. These variables are represented in MCPLOTS by a common identifier, called \texttt{j.m} (MCPLOTS shorthand for jet mass), and are fully described in \texttg{rivet-histograms.map} by the following lines:
    
\medskip
{\ttfamily\small
        \noindent\begin{tabular*}{\linewidth}{@{\extracolsep{\fill}} @{} lllllll @{}}
            \texttt{[beam]} & [proc] & [Ecm] & [par] & [analysis\_histogram] & [obs] & [cuts]\\
            pp & ttbar & 13000 & - & CMS\_2019\_I1764472\_d01-x01-y01 & j.m & cms2019-ttboost \\
            pp & ttbar & 13000 & - & CMS\_2019\_I1764472\_d02-x01-y01 & j.m & cms2019-ttboost \\
        \end{tabular*}
}
    
\medskip
The first three columns here correspond to MC generator initialisation parameters that should be taken from the RIVET analysis page: identities of two incoming beam particles, a hard process, a centre-of-mass collision energy in GeV. The fourth column then lists the kinematic constraints, if necessary. The \texttt{[analysis\_histogram]} column contains the internal RIVET reference from the \texttg{.plot} file. Finally, the \texttt{[obs]} column contains an internal MCPLOTS observable name, and the \texttt{[cuts]} one provides an additional internal label for the analysis (or of individual distributions) that can be used to identify supplementary information that should be added to the plots (see below). If several different analyses include the same observable, the same name is assigned to them. This will cause the corresponding plots to be displayed together, on one and the same page on the MCPLOTS web site, rather than on separate ones. MCPLOTS observables should be named according to the generic scheme \texttt{particle.characteristic}. For example, \texttt{j.m} was used for jet mass above, while, e.g., for a lepton rapidity, we use \texttt{l.y}. These identifiers do not \emph{have} to be standardised; but it is convenient to do so the better to  control and collect how they appear on the web pages. Note that different hard processes are always displayed on separate web pages, so that the distribution of the jet mass \texttt{j.m} from the ttbar process will not be mixed with, say, the \texttt{j.m} from the hard QCD process, for example.
    
The above analysis contains only 2 distributions, but there are RIVET analyses that contain several hundreds of them (for example using different bins for jet energy). In this case, the python script \texttg{parser.py}, located in the \texttg{/scripts/mcprod} folder, can be used to automate the description of the analysis in MCPLOTS. It analyses each \texttt{PLOT} section in the RIVET file and translates it into MCPLOTS format, allowing the user to interactively assign MCPLOTS variables to RIVET distributions, taking into account distributions to which such variables have already been assigned. For the script to work, the user should provide information for the first 3 and the last columns of the \texttg{rivet-histograms.map} file, which are assumed to be unchangeable, the rest of the columns will be filled by the script automatically.
    
All information about the distribution provided in the description line, except kinematic constraints, is propagated to the path to the generated MC distribution, and then to the database, and used to filter and display plots on the website.
    
For some analyses, different running modes are to be specified, for example, when cross sections are measured for e.g. muons and electrons separately and also combined, or for different event selections. An optional machinery of RIVET is handled in MCPLOTS by specifying all possible modes in the configuration file:
    
\medskip
{\ttfamily\small
        \noindent\begin{tabular*}{\linewidth}{@{\extracolsep{\fill}} @{} lllllll @{}}
            \texttt{[beam]} & [proc] & [Ecm] & [par] & [analysis\_histogram] & [obs] & [cuts]\\
            pp & z1j & 13000 & 55 & ATLAS\_2022\_I2077570:LMODE=ELEL\_d01-x01-y01 & z.pt & atlas2022-zj \\
            pp & z1j & 13000 & 55 & ATLAS\_2022\_I2077570:LMODE=MUMU\_d01-x01-y01 & z.pt & atlas2022-zj \\
            pp & z1j & 13000 & 55 & ATLAS\_2022\_I2077570\_d01-x01-y01 & z.pt & atlas2022-zj \\
        \end{tabular*}
}
\medskip
    
\noindent Here, the implementation of the ATLAS analysis \cite{ATLAS:2022nrp} is shown, where cross-section measurements for a Z boson produced with jets are presented. These measurements are performed in the electron and muon channels, and the combination is also provided. This is reflected by the presence of the option \texttt{LMODE=ELEL} or \texttt{MUMU} in the \texttt{[analysis\_histogram]} column, while the absence of any option means the combination. 
    
In this example, kinematic constraints in the \texttt{[par]} column come into play. The MCPLOTS machinery allows to pass phase-space cuts specified in this column to MC generators and thus restrict the hard-process kinematics, to optimise the event-generation efficiency. There are four parameters that can potentially be restricted: the minimum and maximum transverse momentum of the final-state hard-process partons and their minimum and maximum invariant mass, respectively. These parameters are standard and are accepted by almost any MC generator. If these cuts are not necessary the column \texttt{[par]} is filled with the symbol ``\texttt{-}'' as in the previous example. Otherwise they should be provided in the order they were given above, separated by commas, without spaces, and the last unnecessary cuts may be omitted. For example, a cut requiring a minimum hard-process 
$\hat{p}_\perp$ of 25 GeV and a minimum hard-process $\hat{m}$ of 300 GeV, with no maxima, would be specified by the string \texttt{25,-,300}. Such cuts must be applied with extreme care, as discussed in the following subsection.

Note that, since the cross-section decreases very rapidly as the momentum of the leading outgoing parton increases, it is often sufficient to use only the minimum transverse momentum cut. Thus, in the first example above, only the first (minimum hard-process $\hat{p}_\perp$) cut was imposed to ensure an adequate population of histograms. 
    
\subsubsection{MC Hard-Process Phase-Space Cuts}

For many hard processes, in particular ones that contain t-channel singularities and/or s-channel resonances, inclusively generated event samples will be dominated by the singular / resonant structures. This is not always desirable. E.g., if a particular analysis only looks at high-$p_\perp$ tails and/or specific windows in invariant mass, it may be almost impossible to generate a sufficient number of inclusive events (i.e., without any generator-level hard-process cuts) to populate the corresponding phase-space region. Moreover, in some cases, like the aforementioned one of t-channel singularities, generator-level hard-process phase-space cuts (henceforth referred to as generation cuts) are needed a priori to render the physical cross section finite. Not to mention the significant waste of CPU resources it would be to generate event samples for which only a tiny fraction of the events would be used and the rest thrown away. 

The question of finding an optimal set of generation cuts for a given analysis is therefore quite important and has been discussed in the literature, e.g., in the context of V+jet processes~\cite{Alioli:2010qp}. We emphasise that generation cuts refer to a hard partonic process (before showering and hadronisation), and not to the physical particle-level final state, on which analysis cuts are imposed. In general, the generation cuts must be chosen considerably broader (``looser'') than the analysis cuts, as the generation-level hard-process quantities can be shifted both up or down by shower and hadronisation activity. 

\begin{figure}[t]
        \centering
        \resizebox{0.65\textwidth}{!}{
            \includegraphics{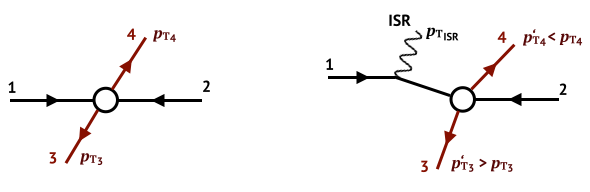}
        }
        \caption{Change of $p_T$ of outgoing particles in the presence of ISR}
        \label{fig:cuts-diag}
\end{figure}
    
For illustration, consider a parton emerging from a hard process. During a parton shower, the transverse momentum of this parton can be both increased or decreased, as illustrated in Fig. \ref{fig:cuts-diag} by partons 3 and 4, respectively, and this effect should be kept in mind when determining a generation cut. The optimal cut must be low enough so as not to lose events that can pass the data-analysis cut after showering and hadronisation and at the same time high enough to ensure that as many as possible of the generated events are in the desired region of phase space, i.e., that as few events as possible are thrown away. In the end, only the speed with which the results are obtained should depend on these cuts, not the final physical distributions themselves. However, there is so far no rigorous automated procedure for how to determine such cuts, and e.g., in the case of  adding new analyses to the MCPLOTS configuration file, it is hardly possible to verify how all of the final physical distributions depend on generation cuts; plotting histograms with sufficiently high statistics is a very resource-consuming task. To address this problem, we have developed two generic methods to determine optimal generation cuts.
    
The first one uses a direct estimate of the number of  events that are ``lost'' (never generated, but should have been) when using the given generation cut. This refers to the number of events that \emph{would not} satisfy the proposed generation cut (i.e., events that would never be generated, with that cut) but which \emph{would} pass the data-analysis cut after showering and hadronisation. For appropriate generation cuts, this number must be zero. Let us describe the procedure in more detail. To start, a proposed initial value of the generation cut, call it $p_{T0}$, is set, for instance only slightly below the given data-analysis cut. Then a test sample is generated and events pass the complete RIVET analysis chain. At this point, the number of events used in the final analysis distributions $N_0$ is obtained. To estimate the number of ``lost'' events, another test sample is generated with the same parameters, but with the transverse momentum of the hard-process final-state partons constrained in a narrow $p_T$ band, $p_{T0}-\Delta<p_T<p_{T0}$, where $\Delta$ is a relatively small value. For the reasons discussed above, some of these events may still pass the data analysis cut, and after passing the full RIVET analysis chain, the number of such events $N_{0,lost}$ is obtained. The latter is a number of events ``lost'' when using the $p_{T0}$ generation cut. If this number is non-zero, a new generation cut value of $p_{T1}<p_{T0}$ is proposed, resulting in a new estimate of the relative number of ``lost'' events $\frac{N_{1,lost}}{N_1}$. The procedure continues until this number becomes negligibly small, and the corresponding generation cut is then considered as an optimal one. In MCPLOTS, the ``loss'' threshold value is set to not exceed a fraction of a percent. 
    
The final result, the value of the optimal generation cut, does not depend on the choice of  $p_T$ bandwidth $\Delta$, other than via the finite resolution caused by iterating in steps of $\Delta$. 

\begin{figure}
    \begin{minipage}[t]{0.48\textwidth}
        \resizebox{1\textwidth}{!}{
            \includegraphics{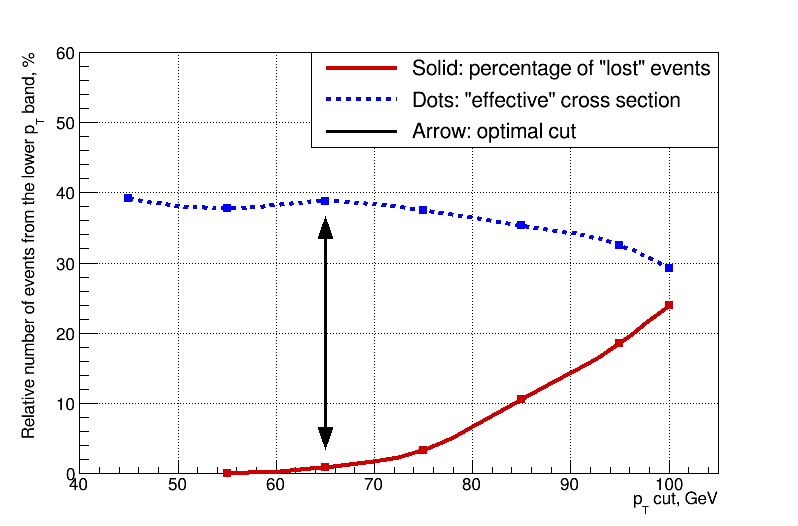}
        }
        \caption{Finding an optimal generation cut using a lower $p_T$ band method (solid red line) and an "effective" cross section method (blue dots in arbitrary units). An optimal cut is represented by the arrow.}
        \label{fig:jetsband}
    \end{minipage}
    \hfill
    \begin{minipage}[t]{0.48\textwidth}
        \resizebox{1\textwidth}{!}{
            \includegraphics{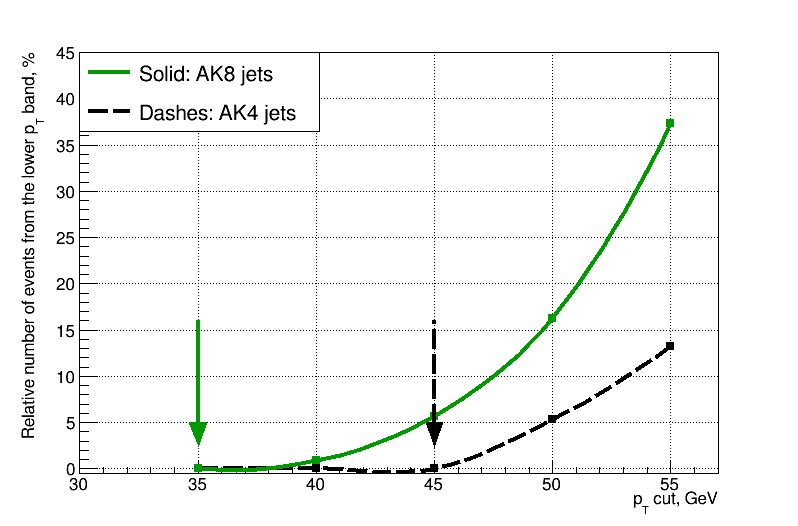}
        }
        \caption{Dependence of the optimal generation cut on the jet clustering algorithm. }
        \label{fig:ak-cuts}
    \end{minipage}
\end{figure}

Some MC generators (and/or process types) do not provide the possibility of setting a constraint on the maximum transverse momentum of the final-state partons, and then the $p_T$ band method cannot be used. In this case, we use a second method. Again, the initial value of the generation cut is set to some (guessed) $p_{T0}$, the test sample is generated and a number of events accepted in the data analysis phase space $N_0$ is obtained, as well as an ``effective'' cross section $\sigma_0\cdot N_0$. If the  $p_{T0}$ generation cut is too high, then lowering it to $p_{T1}<p_{T0}$ will cause the ``effective'' cross section to increase since some of the events that were ``lost'' when using the $p_{T0}$ cut will now be generated and enter in the calculation of the total cross section of accepted events. As long as this cross section increases, the procedure is repeated for a new generation cut $p_{T2}<p_{T1}$. At some point the ``effective'' cross section stabilises and does not change anymore when further reducing the generation cut. The corresponding generation cut is then considered the optimal one. 
    
The \texttt{CMS\_2013\_I1265659} analysis \cite{CMS:2013cvx} in which colour coherence effects are studied in 3-jets events is used to illustrate both methods here. In this data analysis the leading  jet is required to have $p_{T} > 100$ GeV, so several hard QCD samples with different $p_{T}$ cuts from 40 to 100 GeV were generated with Pythia~8. The effect of the $p_{T}$ cut choice is represented in Fig. \ref{fig:jetsband}, where the solid (red) line shows the dependence of the relative number of "lost" events on the cut, the (blue) dots show the dependence of the "effective" cross section (for accepted events) on the cut, and finally the arrow shows the generation cut, which is considered as an optimal one and is used in MCPLOTS. Both methods result in the same value of the optimal generation cut, and the decrease of  the "effective" cross section becomes visible when around 10\% of events are "lost".
    
An optimal generation cut depends not only on the data analysis cut, not only on the kinematics of the process under consideration, but also on the method of reconstruction of physics objects in the event, on the choice of the jet clustering algorithm etc. Thus, the cut will be different for example, when using AK4 and AK8 jets, as illustrated in Fig. \ref{fig:ak-cuts}. The  \texttt{CMS\_2021\_I1920187} analysis \cite{CMS:2021iwu} studying the jets substructure in Z+jet events in various $p_{T, jet}$ bins is used here for the illustration purpose. In Fig. \ref{fig:ak-cuts} the $p_T$ band method is shown for the analysis constraint  $65 \textrm{ GeV} < p_{T, jet} < 88  \textrm{ GeV}$. One can see that the optimal generation cut for AK4 jets is tighter than the one for AK8 jets. In MCPLOTS, the minimal one is imposed to plot distributions with both types of jets.
    
Another important technical note is that a separate generator run is required for each different set of generation cuts. Hence it is useful to choose as small set of different generation cuts as possible. For this purpose, for new analyses implemented in MCPLOTS, the previously imposed $p_T$ cuts are also taken into account, and new sets of cuts only defined when necessary.  

\subsubsection{Displaying the New Plots on MCPLOTS}


For each hard process, each observable defined in the \texttt{[obs]} column of the \texttg{rivet-histograms.map} file is represented on a separate web page. Such a web page contains all distributions of a given observable from corresponding RIVET analyses, with all given kinematic constraints / cuts. While the web pages are generated automatically, the descriptions of hard processes, observables, and their groups are specified in the \texttg{mcplots.conf} file. In fact, the latter defines the correspondence between internal names (as declared in the \texttg{rivet-histograms.map} file) and names that will be displayed on the web page and on plots.

Hard-process names are translated by a line of the form

\medskip
{\ttfamily
    \begin{tabular*}{\linewidth}{l@{ ! } l@{ ! }c }
        process\_name  =  & HTML name & name in LaTeX format \\
    \end{tabular*}
}

\noindent in \texttg{mcplots.conf}. The \texttt{process\_name} is the name of the hard process used in the \texttg{rivet-histograms.map} file, the \texttt{HTML name} column defines the name appeared on the MCPLOTS web site and finally the \texttt{name in LaTeX format} column corresponds to the name appeared on plots as the plotting tool uses LaTeX.

When defining observable names, it is possible to group a set of observables for a better representation. It is done by specifying an optional group name after the equality sign, before the first exclamation mark:

\medskip
{\ttfamily
    \begin{tabular*}{\linewidth}{l@{ ! } l@{ ! }c }
        observable\_name  =  (HTML group name) & HTML name & name in LaTeX format \\
    \end{tabular*}
}

\noindent Again, the \texttt{HTML name} column and the optional \texttt{HTML group name} define the appearance on the web site and the \texttt{name in LaTeX format} is used in the plotting tool.
    
The correspondence between the declaration lines and the website layout is illustrated in Fig.~\ref{fig:plotspage}, which shows the Average Charged Particle Density in the AWAY region variable from the AWAY variable group of the Underlying Event hard process. This variable is described by the following line
    
\medskip
\noindent\texttt{nch-vs-pt-away      =  AWAY ! \&lt;Nch\&gt; vs pT1     ! Average Charged Particle Density (AWAY)}
\medskip
    
Cut names introduced in the column \texttt{[cuts]} have to be translated as well. This is achieved by expanding \texttg{mcplots.conf} with cut declaration lines in the format
    
\medskip
{\ttfamily
        \begin{tabular*}{\linewidth}{l@{ ! } l@{ ! }c }
            cut\_name  =  & HTML name & name in LaTeX format \\
        \end{tabular*}
}

These names will appear in $\chi^2$ tables, as illustrated in Fig.\ref{fig:typdet}. The first two lines are the result of adding the following to the configuration file:
    
\medskip
{\ttfamily
        \begin{tabular*}{\linewidth}{l@{ ! } l@{ ! }l }
            cms-coh        = & CMS 2013 (Forward) & 0.8<|\#eta\_\{2\}|<2.5 \\
            cms-coh-ctr    = & CMS 2013 (Central) & |\#eta\_\{2\}|<0.8
            
        \end{tabular*}
}

 \subsection{Adding a New MC Generator to MCPLOTS}

Since Test4Theory and MCPLOTS are intended to  be almost fully automated, with minimal overhead of additional human manpower, we do not maintain our own build procedures for MC generators but instead rely on global installations of MC binaries provided by the CERN EP-SFT group. Specifically, Test4Theory and MCPLOTS are exclusively based on the LHC Computing Grid (LCG) software trees that are available on the CernVM File System (CVMFS)~\cite{Blomer:2012cm} (or on AFS for earlier versions). The first prerequsite for including a new MC generator on MCPLOTS is therefore that it must be included in an LCG release. Inclusion of generators not installed at CERN is not supported.

Adding a new MC generator to the project thus requires the following steps: 
\begin{enumerate}
    \item Specify the LCG installation location of the generator;
    \item Specify the commands required to run this generator in Test4Theory;
    \item Add the generator (and any non-default tunes of it) for display on  MCPLOTS. 
\end{enumerate}

On CVMFS, generators are normally installed in the directory  
\begin{center}
\texttg{/cvmfs/sft.cern.ch/lcg/releases/LCG\_*/MCGenerators}, 
\end{center}
where "*" is the \texttt{LCG} installation version number. We now illustrate the implementation process of a new generator by the example of adding the POWHEG-BOX generator, located in
    
\medskip
\noindent\texttt{/cvmfs/sft.cern.ch/lcg/releases/LCG\_96/MCGenerators/powheg-box-v2/r3043.lhcb}
\medskip
    
\noindent where \texttt{r3043} means the revision 3043 of POWHEG-BOX version 2. Each installation is available for multiple platforms, and it is not necessary to choose one; this will be done automatically by MCPLOTS, according to the execution environment. After the generator is located, it is necessary to specify the path to it in the file \texttg{locations.map} which can be found in the usual \texttg{configuration} directory. In this file the data is organised in columns. The first one is a generator name, the second one contains its version, then the LCG release number (name), the fourth optional column is an actual path to the generator version if it doesn't fully correspond to the first and second columns, and finally there are extra optional parameters for generator runs. The above-mentioned POWHEG-BOX installation is described as follows:
    
\medskip
    
{\ttfamily
        \noindent\begin{tabular*}{\linewidth}{@{\extracolsep{\fill}} @{} lllll @{}}
            powheg-box & r3043 &  LCG\_96 & powheg-box-v2/r3043.lhcb & pythia8=244 \\
        \end{tabular*}
}
        
\noindent Here the last parameter is the Pythia~8 version to be used for the parton-shower generation.
    
After adding the description line for the new generator, code must be added to actually run  it within\\ Test4Theory/MCPLOTS. This is done by the \texttg{rungen.sh} script, located in the directory \texttg{scripts/mcprod}. Each generator is started in the \texttt{run()} function, which must be extended with commands to start a new one. Although these commands naturally differ between each generator, two basic blocks common to all cases can be distinguished. 
    
The first one is the creation of a temporary steering file to start
the generator. This file contains information from the template of a
steering file, which must be prepared in advance for each hard process
and contains the necessary common parameters for starting the
generation of MC events, and optionally from a tune file, from which
model parameters are propagated. Templates of generator steering files
are located in the \texttg{configuration} directory and are named
\texttg{<generator name>-<hard process>.params}. They contain
placeholders for common parameters necessary to start the generation
of MC events for a given hard process: beam characteristics, number of
events to be generated, hard-process-specific parameters and so on
according to the generator capabilities. These placeholders are to be
substituted by command line parameter values from the run
command. MCPLOTS machinery allows the use of matching/merging of the
matrix element generator results and the parton shower. This way, for
example, the result of POWHEG-BOX is merged with a parton shower in
Pythia~8, and in this case the steering file template is divided into 2 sections each starting with a keyword of the form \texttt{[generator-name]}. Each of these sections contains a set of parameters needed to run the corresponding generator: \texttt{[powheg]} to start the generation of events and \texttt{[pythia8]} to start the parton shower.
    
The second block of commands is used to start the generator with the steering file created in the first block, and each specific generator should be started according to its own guidelines.
    
Note, that environment setup might be necessary for the selected generator installation: setup of the proper \texttt{gcc} compiler version, \texttt{LHAPDF} version etc. The \texttg{set\_environment\_LCG\_*} functions in the same \texttg{rungen.sh} file are used for this purpose. 
    
\subsubsection{Adding a New Generator Version to MCPLOTS}
     
Often starting a new version of the generator does not differ from starting an already added version. Then it is enough to add the path to the new version to the file \texttt{locations.map}.  However, if the installation of the new version uses a different version of \texttt{LCG}, it may be necessary to add a new environment setup function. For example, adding the next version of the POWHEG-BOX generator 
     
\medskip
     
{\ttfamily
         \noindent\begin{tabular*}{\linewidth}{@{\extracolsep{\fill}} @{} lllll @{}}
             powheg-box & r3744 &  LCG\_104 & powheg-box-v2/r3744.lhcb3.rdynamic & pythia8=309 \\
         \end{tabular*}
}
     
\noindent entailed adding an environment setup function for \texttt{LCG\_104}.
     
However, if the new version requires specific commands, it will be necessary to make corresponding changes to the generator launch code in the \texttg{rungen.sh} script.
    
\subsection{Adding New Tunes to MCPLOTS}
      
Adding a generator tune in most cases means creating a configuration file with the name \texttg{generator-tunename.tune} containing a set of model parameters that is used when starting this generator. This set of parameters is to be copied to the steering file, which is created to start the generator in the script \texttg{rungen.sh}. For example, to study the hadronic rescattering model, the corresponding rescattering tune is used for the Pythia~8 generator.  The tune file \texttg{pythia8-rescatter.tune} consists of lines containing parameters switching on hadronic rescattering:
       
\medskip
\noindent\texttt{HadronLevel:Rescatter = on\\
          Fragmentation:setVertices = on\\
          PartonVertex:setVertex = on
      }
\medskip
      
Information about the tune should be placed in the file \texttg{generator-tunes.map}, which generally defines in which versions of the generator the tune can work, as well as possible additional parameters that depend on the generator. The rescattering tune is described in the file \texttg{pythia8-tunes.map} by the following lines:
      
\medskip
      \noindent\texttt{8.303 ee rescatter 0 \\
      8.303 pp rescatter 0 
    }
\medskip
    
\noindent which mean that the tune works in version \texttt{8.303} (and later) for electron and proton beams, and that it is used together with the \texttt{default} tune encoded by the internal Pythia~8 parameter \texttt{0}.  

\section{Updating the MCPLOTS Website}
\label{sec:update}
    
The previous Section explains how to implement a new analysis, generator (version) or tune to the MCPLOTS project. The next natural step is to prepare new MC distributions, make them public on the website and update the database accordingly.
    
Once new analyses, generators or tunes have been added, the new MC distributions can be produced for these new settings. This can be done through volunteer computing or manually. The first method is normally used for the official production to update all distributions on the main MCPLOTS website, and the second one, manual small scale production, can be very useful for debugging purposes. Distributions are generated by running the script \texttg{runRIVET.sh}, which is located in the directory \texttg{scripts/mcprod}. Command to run the script is
    
\medskip
\noindent\texttt{./runRIVET.sh [mode] [beam] [process] [energy] [params] [specific] [generator] [version] [tune] [nevts] [seed]}\\
    
\noindent where all arguments are mandatory and mean the following:
\begin{itemize}
        \item \texttt{mode}: parameter necessary to set the environment properly. It should be set to \textit{local} value, other values should be used when running this script from a wrapper like \texttt{runAll.sh};
        \item \texttt{beam, process, energy, params}: MC generator initialisation parameters (identities of two incoming beam particles, a hard process, a centre-of-mass collision energy in GeV and kinematic constraints) used in \\ \texttg{rivet-histograms.map} file and described in Sec. \Ref{sec:dev};
        \item \texttt{specific}: generator-specific settings, for example, jet bins for the \textsc{Alpgen} generator;
        \item \texttt{generator, version, tune}: the generator name, its version and its tune to run;
        \item \texttt{nevts}: number of events to generate;
        \item \texttt{seed}: initial seed for the random number generator.
\end{itemize}
    
When this script is run, all distributions from the \texttg{rivet-histograms.map} file that meet the specified parameters will be selected and generated, the results will be put in the \texttg{dat} folder. To check distributions from a particular data analysis, it is useful to set the \texttg{ANALYSIS} environment variable to run the script. The project also allows you to use more variables to control the script, such as  \texttg{MKHTML} and  \texttg{HTMLDIR}, which are responsible for creating an HTML file with all produced histograms and setting the path to this file. For example, to generate and be able to immediately look at distributions from \texttt{CMS\_2019\_I1764472} analysis with the generator Pythia~8 version \texttt{244}, \texttt{default} tune, with 1000 events, run the \texttg{runRIVET.sh} script as follows:
    
\medskip
\noindent\texttt{ANALYSIS=CMS\_2019\_I1764472 MKHTML=1 ./runRIVET.sh local pp ttbar 13000 - - pythia8 \\8.244 default 1000 1}\\
    
\noindent There are no kinematic constraints and generator-specific settings, so the corresponding mandatory parameters are replaced by \texttt{"-"}. Note that the given number of events means the number of generated events in the whole given phase space, and does not take into account the data analysis event selection.
    
The \texttg{runRIVET.sh} script can only be used to run a single configuration. Here by configuration is meant a combination of generator-version-tune-hard process-beam energy-cuts. To perform multiple runs the script \texttg{runAll.sh} should be used. Command to execute it:
    
\medskip
\noindent\texttt{./runAll.sh [mode] [nevt] \{filter\} \{duration\}}
    
\noindent where the two first arguments are mandatory and the two latter ones are optional, with the following meanings:
    
\begin{itemize}
        \item \texttt{mode}: Parameter governing the distribution of MC generator runs. If mode is set to \textit{list}, the script simply returns a list of all possible generator runs. \textit{local} will queue all desired runs on the current desktop, while \textit{lxbatch} distributes jobs on the lxplus cluster at CERN.  Note, that \textit{lxbatch} service is available only from lxplus machines;
        \item \texttt{nevt}: number of MC events per run;
        \item \texttt{filter}: regular expression string used to filter all possible MC configurations. For instance, if setting as a string a configuration that contains all the parameters for running the \texttg{runRIVET.sh} script, then only this one configuration will be run. Another example: to get the list of all possible \texttt{ee} runs of generator Pythia~6 with default tune, the following command should be executed\\
        \noindent\texttg{./runAll.sh list 1M "ee .* pythia6 .* default"};
        \item \texttt{duration}: limit on run time duration (optional, default is 5000 s). This parameter should be changed if jobs take more time to complete (for example in the case of a large number of events).
\end{itemize}
    
\noindent This script executes the script \texttg{runRIVET.sh} for all the desired input settings. The resulting distributions will be placed in the \texttg{dat} directory.
    
To display the results of new MC runs, it is necessary to update the database with the new histograms. These new distributions should be added to the \texttg{www/dat} directory on the server, and then the \texttg{updatedb.sh} script, located in the \texttg{www} directory, should be run. It scans the contents of the \texttg{dat} directory, and automatically updates the database according to the changes in it.
    
The final step of the site update is the calculation of $\chi^2$ values for all MC distributions. To do this, it is necessary to open the \texttg{chi2calc.php} page in a web browser and run the calculation. This will delete all existing files that store information about these values, $\chi^2$s will be recalculated for all distributions from the \texttg{dat} directory, according to the information from the updated database.

\section{Conclusions}
Event-generator validation is evolving quickly. New methodologies and tools are appearing in step with increasingly comprehensive and sophisticated physics models and experimental constraints. Higher precision targets, e.g.\ at the LHC, imply requirements of better understanding and charting of modelling uncertainties. Event-generator models --- and tuning --- are also finding wider applications, e.g., in heavy-ion physics, cosmic-ray physics, neutrino physics, and at B factories.

The HepDATA measurement archive~\cite{Whalley:1989mt,Buckley:2006np,Maguire:2017ypu} and the RIVET analysis preservation tool~\cite{Waugh:2006ip,Buckley:2010ar,Bierlich:2019rhm} define the current state of the art in ensuring that modern (and future) event-generator models can be compared, consistently, to both new and legacy measurements. Some effort --- and time --- is still required to do this, and hence instant feedback is rarely possible. 

MCPLOTS is intended as a ``library of congress'' providing up-to-date and comprehensive comparisons for a wide set of event-generator models. These can quickly be consulted in a few clicks, and both plots and run-card setups appropriate to each generator and analysis are made available for download. So far, the site mainly covers high-energy ee and pp collider measurements, this being the core research area of its main developers and probably also the most extensively represented in terms of MC models and RIVET analyses, but there is no in-principle reason why this could not be expanded in the future.

\begin{figure}[tp]
    \centering
    \resizebox{0.99\textwidth}{!}{
    \includegraphics{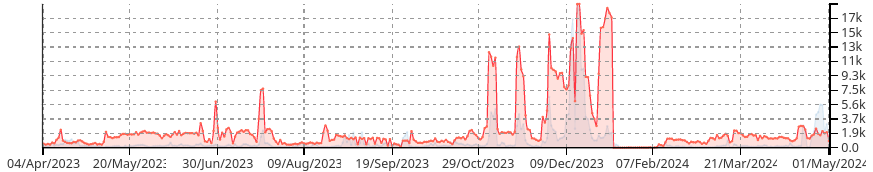}
    }
    \caption{The history of unique visitors per day of MCPLOTS website over the last year. The high spikes corresponds to robots activity, the average non-robot visits rates is 1000 per day.\label{fig:visitors}}
\end{figure}

The project has been evolving since about 2010. Computing power is donated by volunteers via the LHC@home Test4Theory project, which pioneered virtualisation technology for volunteer clouds. 
The first full-fledged public version of the site was documented in 2013~\cite{Karneyeu:2013aha}. The website is in the regular use (Fig.~\ref{fig:visitors}).

In the current paper, we have summarised comprehensive modernisations and technical updates that have been carried out over the last few years. The front end has been given a makeover with a more modern layout, and the site and underlying functionality have been streamlined. A large number of further RIVET analyses have also been added, with corresponding run cards, etc. Efforts have also been made to facilitate further community-based extensions and developments, which we envision as a viable model for sustainable future development of the project, e.g., in some of the directions not  covered by the  authors of this document. To this end, the project is made available via CERN Gitlab: 
\begin{center}
\texttg{https://gitlab.cern.ch/MCPLOTS/mcplots}~.
\end{center}

\paragraph{Data Availability Statement}
All data used by the MCPLOTS project are
available at: \url{https://mcplots.cern.ch}. The source code for MCPLOTS is
available at: \url{https://gitlab.cern.ch/MCPLOTS/mcplots}.

\paragraph{Acknowledgements}
We are grateful for substantial support for this work from the LHC Physics Centre at CERN (LPCC). PS is supported by the Australian Research Council via Discovery Project grant DP220103512 ``Tackling the computational bottleneck in precision particle physics'' and by the Royal Society Wolfson Visiting Fellowship ``Piercing the precision barrier in high-energy particle physics''. 

\bibliographystyle{JHEP}
\bibliography{mainlib}

\end{document}